\documentclass[table,dvipsnames]{optica-article}
\journal{opticajournal}
\articletype{Research Article}
\usepackage{enumitem}
\usepackage{multirow}
\usepackage{hyperref}
\usepackage{cite}
\usepackage{upgreek}
\usepackage{babel}
\hypersetup{
    colorlinks=true,
    linkcolor=blue,
    citecolor=cyan,
    urlcolor=green,
}

\usepackage{tikz,xcolor,hyperref}
\hypersetup{colorlinks=true,linkcolor=blue,urlcolor=blue,citecolor=blue,pdfstartview={XYZ null null 1.00},pdfauthor={Tamás Csizmadia}}
\definecolor{lime}{HTML}{A6CE39}
\DeclareRobustCommand{\orcidicon}{%
	\begin{tikzpicture}
	\draw[lime, fill=lime] (0,0) 
	circle [radius=0.16] 
	node[white] {{\fontfamily{qag}\selectfont \tiny ID}};
	\draw[white, fill=white] (-0.0625,0.095) 
	circle [radius=0.007];
	\end{tikzpicture}
	\hspace{-2mm}
}

\foreach \x in {A, ..., Z}{%
	\expandafter\xdef\csname orcid\x\endcsname{\noexpand\href{https://orcid.org/\csname orcidauthor\x\endcsname}{\noexpand\orcidicon}}
}

\newcommand{\RNum}[1]{\uppercase\expandafter{\romannumeral #1\relax}}

\begin{document}
\title{Active stabilization for ultralong acquisitions in an attosecond pump--probe beamline}

\author{Tam\'{a}s Csizmadia\authormark{1,*}\orcidC{}, L\'{e}n\'{a}rd Guly\'{a}s Oldal\authormark{1}\orcidG{}, Barnab\'{a}s Gilicze\authormark{1}\orcidL{}, D\'{a}niel S\'{a}ndor Kiss\authormark{1}, Tam\'{a}s Bartyik\authormark{1}\orcidS{}, Katalin Varj\'{u}\authormark{1,2}\orcidR{}, Subhendu Kahaly\authormark{1}\orcidA{}, Bal\'{a}zs Major \authormark{1,2}\orcidB{}}
\address{\authormark{1}ELI ALPS, ELI-HU Non-Profit Ltd., Wolfgang Sandner utca 3, H-6728 Szeged, Hungary\\
\authormark{2}Institute of Physics, University of Szeged, D\'{o}m t\'{e}r 9, H-6720 Szeged, Hungary\\}
\email{\authormark{*}\textcolor{black}{\normalfont{E-mail:}}~\href{tamas.csizmadia@eli-alps.hu}{tamas.csizmadia@eli-alps.hu}}

\begin{abstract}
Attosecond time-resolution experiments using noncollinear interferometers require precise and active control of the optical delay to prevent instabilities---including both slow drifts and rapid vibrations---that can obscure the time evolution of the physical system under investigation. In this work, we present the design and results of stability measurements for a double interferometer setup for extreme ultraviolet-infrared pump--probe spectroscopy. The~attosecond pump--probe setup is driven by a high-average-power, high-repetition-rate laser system and offers sub-optical-cycle ($\pm 81~\emph{as}$) stability with a fast feedback rate over extended periods (up to several days). Due to the noncollinear arrangement, the setup enables independent control of both amplitude and phase in the two arms even across significantly different spectral regions. As a proof of concept, we demonstrate attosecond beating in angle-resolved photoemission during two-photon, two-color photoionization, highlighting the broad potential of the system for kinematically and dynamically complete studies of atomic-scale light--matter interactions.
\end{abstract}

\section{Introduction}\label{sect_introduction}
The investigation of physical processes with attosecond (\emph{as}) time-resolution provides a glimpse into the realm of complex quantum interactions. Since the initial demonstration of time-resolved measurements with such resolution over two decades ago \cite{Hentschel2001, Itatani2002}, experimental techniques have flourished into diverse applications in the study of ultrafast electron dynamics across atomic \cite{Alexandridi2021}, molecular \cite{Ahmadi2022} and condensed matter systems \cite{Vogelsang2024}.\par
Pump--probe spectroscopy is the primary technique used to achieve such high temporal resolution. In this method, an initial light pulse, referred to as the pump, triggers transient dynamics in the studied system, which is then probed by a second light pulse delayed by a controlled interval. By varying the time delay between the two pulses, the full temporal evolution of the system can be captured through stroboscopic images of the dynamics.\par
In attosecond resolution pump--probe measurements, the interferometer can be configured in either a noncollinear or in a collinear arrangement depending on the experimental needs \cite{Mandal2021}. Collinear interferometers are  particularly valued for their compactness, stability, and ease of alignment. Their inherent stability and matched wavefronts have enabled technological and scientific milestones, such as the first generation of isolated attosecond pulses from a single laser cycle \cite{Goulielmakis2008} and 12-\emph{as} stability during extended pump--probe experiments \cite{Ertel2023}.\par
On the other hand, noncollinear interferometers---such as those realized in Michelson, Mach-Zehnder, or Sagnac configurations---offer increased versatility by enabling precise control over the beam characteristics in the two arms independently, as well as accommodating larger pump--probe delays. The ability to independently adjust amplitude, phase, or polarization across different spectral regimes is crucial for a wide range of applications. These include the complete characterization of the time evolution and polarization state of attosecond pulses \cite{Han2023}, the generation of circularly polarized attosecond pulses \cite{Hickstein2015} and ultraviolet--extreme ultraviolet (UV--XUV) time-resolved photoemission spectroscopy \cite{Crego2024}, among others. Many applications also benefit from delays longer than a few optical cycles, for example the study of photoabsorption-induced charge transfer \cite{Murillo-Sánchez2021}, time-resolved electronic and nuclear motion \cite{Xie2014} and relaxation processes \cite{Lucchini2022,Inzani2025,Schiller2025}.\par
The high sensitivity of noncollinear interferometers demands not only careful alignment, but also active stabilization of the optical path difference.
This is especially relevant for long Mach--Zehnder setups, which may involve potentially multiple recombination points for parallel experimentation \cite{Csizmadia2023} , as well as in systems driven by high average power lasers \cite{Ye2022}, where thermal loads on optical components can lead to unintended, simultaneous changes in beam pointing and optical delay. Another challenge arises, when the interfering pulses have significantly different central wavelengths, which can prevent optical coherence---a common issue for many XUV--infrared (IR) pump--probe beamlines \cite{Shirozhan2024}. While interferometric photoelectron measurements can provide optical delay information and offer \textit{in situ} stabilization \cite{Luttmann2021}, the feedback rate in this case is slower than optical methods due to the limitations imposed by the dead time of data acquisition electronics, space charge effects, or the violation of particle coincidences.\par
To achieve faster feedback rates, Fourier-transform spectral interferometry (FTSI) \cite{Lepetit1995} can be employed, often with the addition of a continuous wave (cw) laser beam directed through optics separate from those used in the XUV--IR interferometer \cite{Chini2009}. While this noncoaxial approach can effectively stabilize vibrations of mechanical origin, it cannot fully address instabilities caused by heat load, which affect optics that are in direct contact with the high power beams. One solution is to sample a portion of the pump radiation before the XUV shaping and steering optics \cite{Srinivas2022} or reroute it around them \cite{Schlaepfer2019,Luo2023}, ultimately recombining the pump and probe beams. A coaxial cw interferometer, as demonstrated by Vaughan et al., involves a technically more complex setup, incorporating a second cw laser and an additional feedback loop to stabilize a supplementary optical arrangement that ensures the collinearity of the interfering cw beams before detection \cite{Vaughan2019}.\par
In this work, we demonstrate a method to stabilize a versatile noncollinear XUV--IR interferometer  using an auxiliary interferometer with the aim of conducting time-resolved coincidence spectroscopy, which inherently requires interferometric stability during long-term (hours to days) acquisition. As a proof-of-principle experiment, we investigate the two-photon two-color ionization of argon gas using the Reconstruction of Attosecond Beating By Interference of Two-photon Transitions (RABBITT) technique \cite{Paul2001,Mairesse2003}. Through this approach, we extract the angular dependence of the sideband (SB) phases, providing an additional, crucial control parameter of the ionization process.\par

\section{Experimental setup}\label{sect_experimental_setup}
\subsection{XUV--IR interferometer}
\begin{figure*}
  \centering
  \includegraphics[trim=0.4cm 15.4cm 0.1cm 9.9cm,clip,width=1.0\linewidth]{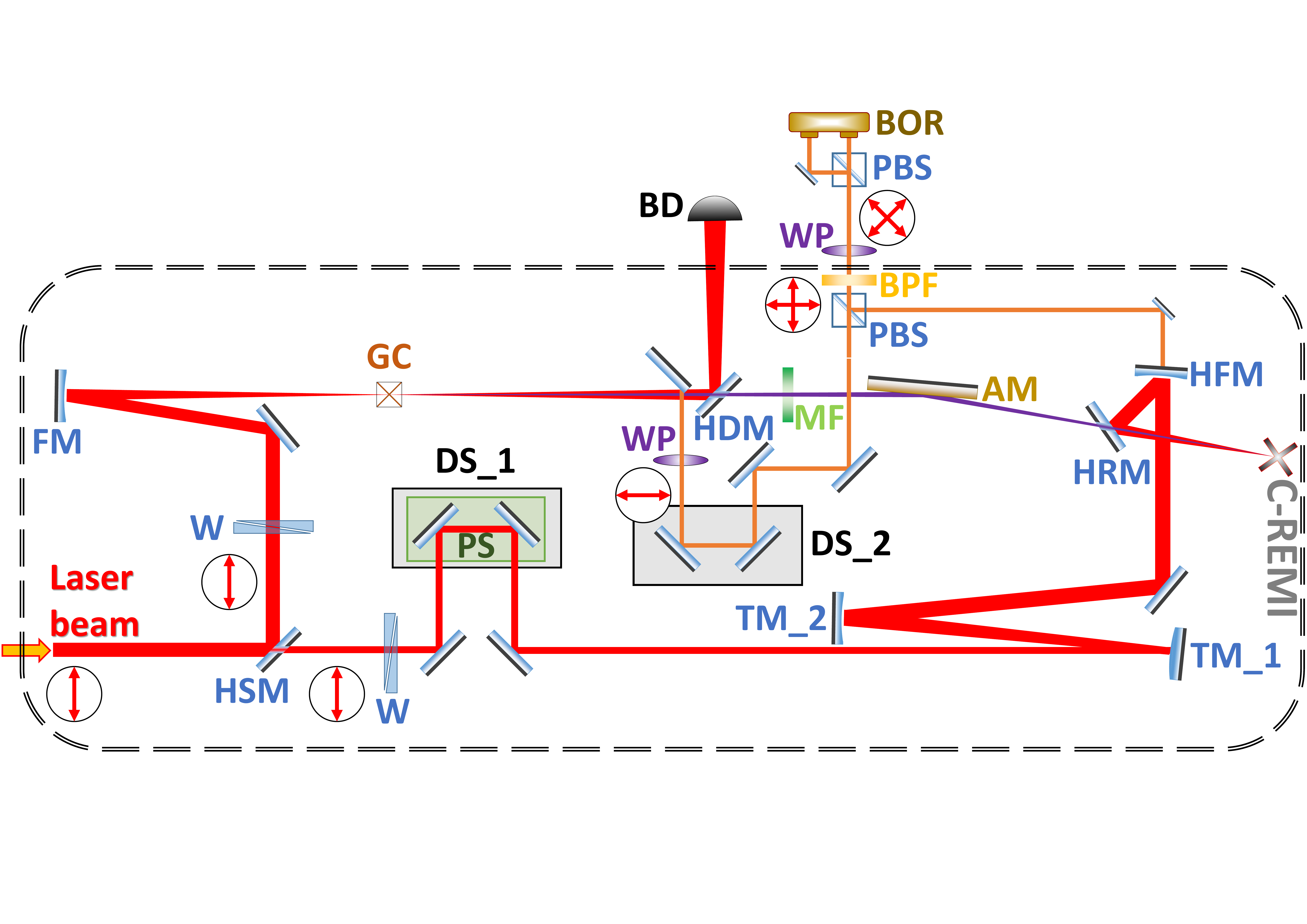}
  \caption{Schematic optical layout of the main XUV (purple) and IR (red) interferometer complemented with the auxiliary interferometer (orange beam path). HSM: holey splitting mirror; W: wedge pair; FM: focusing mirror; GC: gas cell; HDM: holey dumping mirror; BD: beam dump; MF: metallic filter; AM: aspherical mirror; DS\_1: main delay stage; PS: piezo stage; TM\_1 \& TM\_2: telescope mirrors; HFM: holey focusing mirror; HRM: holey recombination mirror; WP: waveplate; DS\_2 auxiliary delay stage; PBS: polarizing beam splitter; BPF: bandpass filter; BOR: balanced optical receiver. The red arrow(s) mark the direction of polarization(s) in the given section. The in-vacuum part of the setup is circumscribed with a dashed solid line.}\label{Fig_layout}
\end{figure*}

The stabilization setup is implemented at the HR Gas beamline \cite{Ye2020,Ye2022,Shirozhan2024} of ELI ALPS \cite{Kuehn2017,Charalambidis2017}. The experiment was driven by the 10~kHz repetition rate, 10~W average power HR Alignment laser system, delivering 35~fs pulses centered at 1030 nm. The laser system is a combination of an Yb:KGW frontend and two nonlinear compression stages complemented by a set of chirped mirrors \cite{Gilicze2025}. Essential components of the optical setup is presented in Fig.~\ref{Fig_layout}. Inside the beamline, the laser beam is split and sent into two, approximately 15~m long, arms of a Mach-Zehnder interferometer (referred to as "main interferometer") using a holey splitting mirror (HSM) for generating high-order harmonic radiation and probing time-resolved experiments, respectively. A~wedge pair (W) is installed in both arms for spectral phase tuning. The pump beam, which contains about 80\% of the pulse energy in an annular beam shape, is focused into a custom-designed patented gas cell assembly (GC) \cite{Filus2022} using a focusing mirror (FM, focal length: 1000~mm) to produce XUV radiation. After generation, the annular beam is dumped by a holey mirror (HDM). Various filters (MF) can be used to select a specific spectral region, compensate for the inherent chirp of harmonics \cite{Varju2005}, or remove the small portion of diffracted infrared beam copropagating with the XUV radiation. The XUV light is finally focused into a reaction microscope (also known as Cold Target Recoil Ion Momentum Spectrometer, hereafter referred to as C-ReMi \cite{Dorner2000,Ullrich2003,Schmidt-Böcking2021}) using a set of grazing incidence XUV optics, simplified into a single aspherical mirror (AM) for clarity. The central part, functioning as the probe beam, is directed onto a retroreflective mirror pair mounted on a translational delay stage (DS\_1) that controls the pump--probe delay. The beam is subsequently expanded by a telescope (TM1 and TM2) and reflected by two holey mirrors (HFM and HRM) before getting recombined with the harmonic radiation.\par
\subsection{Auxiliary interferometer}
For the auxiliary interferometer used in stabilization, a weak portion of the near-infrared beam is sampled from both arms of the main interferometer, in a manner that ensures that the steering optics, which are under high thermal load, are shared by both the XUV--IR and auxiliary optical layout. The working principle of the auxiliary interferometer is based on a polarization-optical method using homodyne detection \cite{Hu2015}. The setup consists of three main subsystems dedicated to optical shaping, detection, and electronic signal processing, including a feedback-based proportional–integral–derivative (PID) control loop.\par
In the optical shaping part, orthogonal polarization is adjusted between the two linearly polarized arms using a halfwave plate (WP). After collinear recombination, the temporal coherence length between the two arms is increased with a bandpass filter (BPF) for the robustness of the day-to-day alignment of the setup. The polarization of the overlapping beams is further rotated by 45$^\circ$ using a second WP to maximize sensitivity for detecting delay-dependent polarization change in the $s$~and~$p$ polarization directions. In the second subsystem, a polarizing beam splitter (PBS) cube splits the beam into two orthogonally polarized parts, between which polarization-dependent differential intensity is measured using a balanced optical receiver (BOR) \cite{Huppert2015}. The voltage on the differential output of the BOR is then digitized and the digital signal is fed into a software feedback loop controlling a fast, high-precision piezoelectric stage (PS). During pump--probe experiments, a third positioner (DS\_2)---located in the auxiliary setup---counteracts the controlled changes of the XUV--IR delay maintaining the locking point close to the maximum sensitivity, thereby reducing cyclic errors.
\begin{figure*}
  \centering
  \includegraphics[trim=1.5cm 0cm 4.0cm 0cm,clip,width=1.0\linewidth]{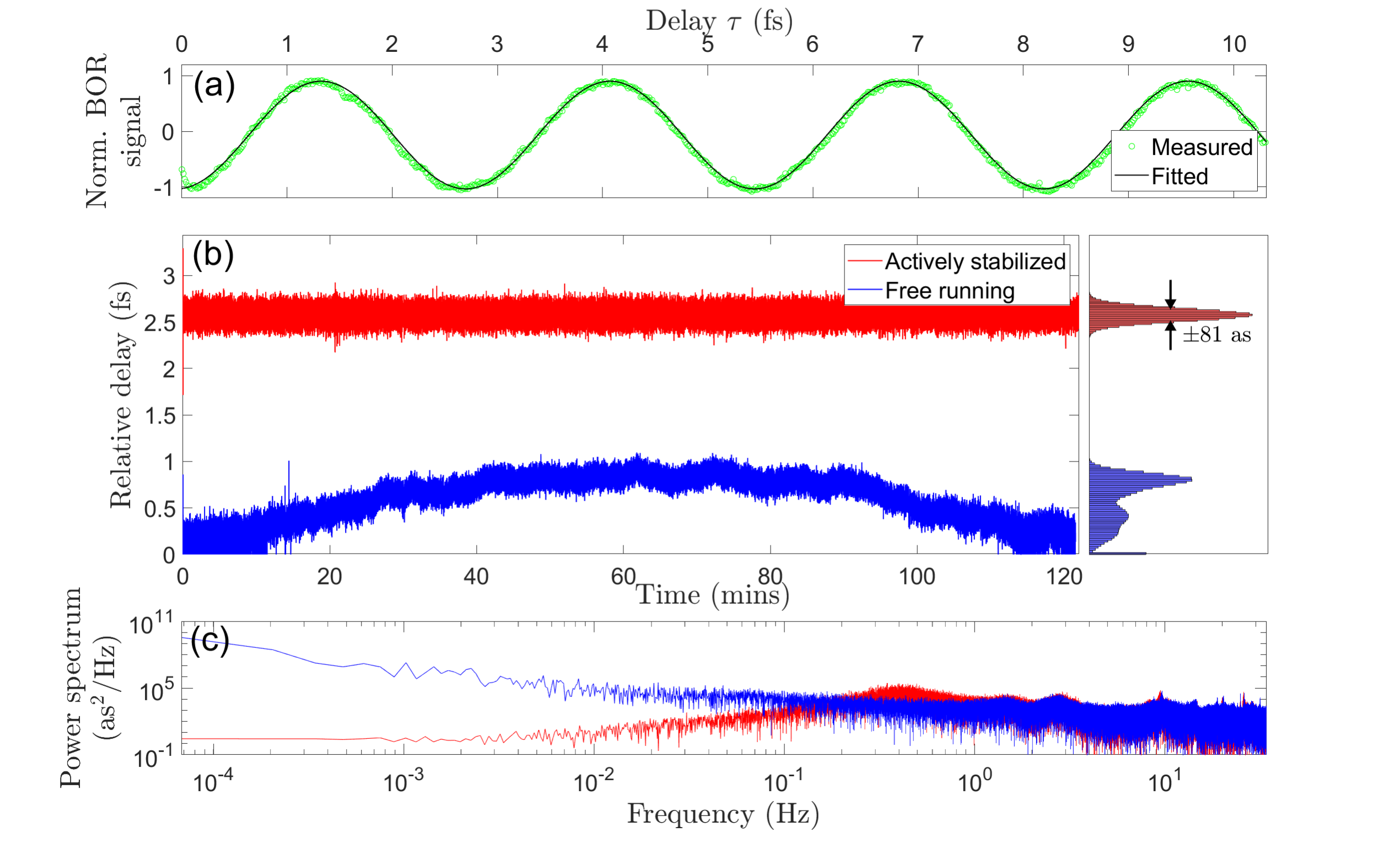}
  \caption{Calibration and temporal behavior of the auxiliary interferometer. (a):~Normalized differential signal of the BOR (green circles) as a function of relative IR-IR delay used to calibrate the feedback-loop (black line). (b):~Long-term evolution and distribution of the relative delay in the auxiliary interferometer with (red) and without (blue) active feedback. (c): Fourier transform of the phase fluctuations.}
  \label{Fig_AuxPD}
\end{figure*}

\section{Results}\label{sect_results}
\subsection{Subcycle and multicycle stability}\label{subsect_stab}
The natural hysteresis and repeatability of the piezoelectric translation stages, accompanied by day-to-day variations in the optical alignment necessitates calibration before initiating the feedback loop (Fig.~\ref{Fig_AuxPD}(a)). During this process, the stabilizing stage is moved with a constant speed to parametrize the function between the measured differential voltage of the photodiode pair $U_{\mathrm{BOR}}$ and the relative optical delay $\tau$:
\begin{equation}
   U_{\mathrm{BOR}}=A\mathrm{cos}(B\tau+C)+D,\label{Eq_Calib}
\end{equation}
where $A$, $B$, $C$ and $D$ are fitting coefficients. It is assumed that the relationship between the relative optical delay and the position of the piezoelectric stabilizing stage remains linear over the short (few seconds) duration of the calibration run. The validity of this assumption places a lower limit on the achievable stability, and can be estimated as the horizontal full width at half maximum (FWHM) fitting error of the relative optical delay, which is $\pm13~\emph{as}$. The relative change in the optical delay is calculated from the BOR signal using Eq.~\eqref{Eq_Calib} by our computer script. This script also calculates the corresponding error signal and adjusts the position of the piezoelectric translation stage based on PID control. The control parameters were tuned using the Ziegler–Nichols method \cite{ZieglerNichols1993}. Figure~\ref{Fig_AuxPD}(b) shows the evolution of the relative delay in the auxiliary interferometer over a two-hour period. When the feedback control is not activated (blue line), an uneven drift is visible in the interferometer with the maximum slope of around 850~\emph{as} per hour. During active stabilization, this drift is essentially eliminated with a residual FWHM error of $\pm81~\emph{as}$. The PID loop operates at an iteration rate of about 70~Hz, which is sufficient to compensate for disturbances that are order of magnitudes faster than the scale of typical data collection times in photoelectron spectroscopy, as shown by the efficient damping of slow fluctuations in Fig.~\ref{Fig_AuxPD}(c). Even though the computer-based PID loop provides high level of flexibility for calibration and tuning the control loop, it is worth noting that, as a future upgrade, using an analog variant combined with already existing hardware components would allow for notably faster loop speed, up to shot-to-shot stabilization.\par

\begin{figure*}
  \centering
  \includegraphics[trim=0.0cm 0.0cm 0.0cm 0cm,clip,width=1.0\linewidth]{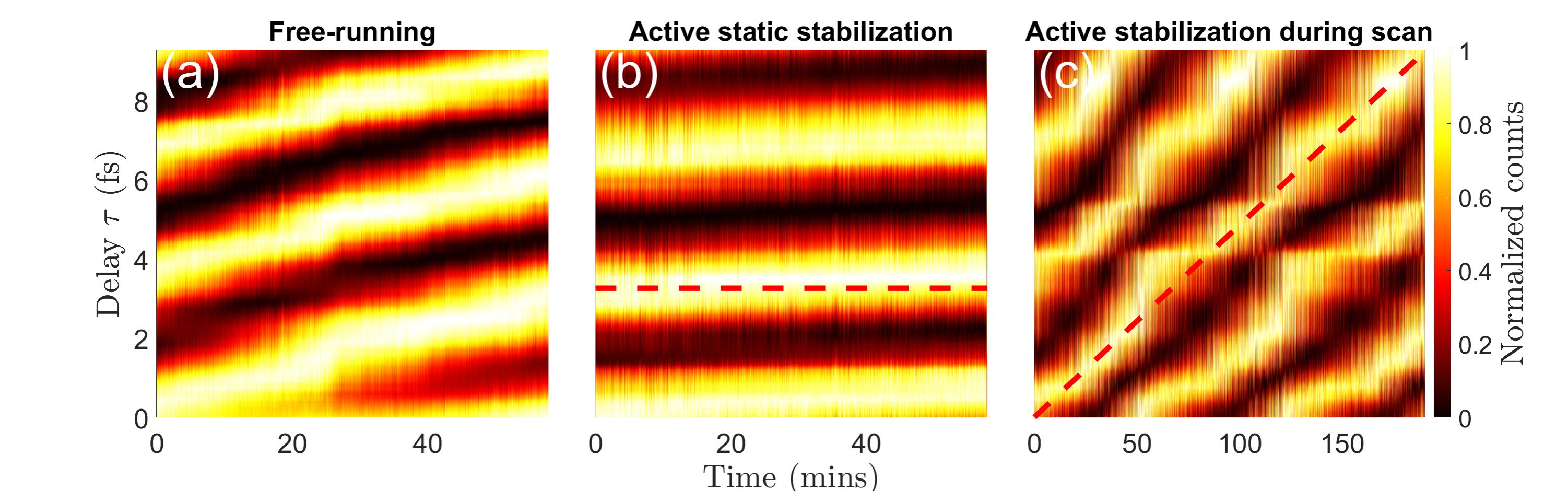}
  \caption{Variation of interference fringes in the main interferometer, when the active stabilization in the auxiliary interferometer is (a): inactive, (b): active during a static acquisition and (c): active during a long linear scan of the optical delay. The relative delay was adjusted to values marked with the dashed red line.}
  \label{Fig_MainFringes}
\end{figure*}
Due to the slight difference in the optical path of the two interferometers, it is important to check and validate the behavior of the XUV--IR interferometer, when the optical delay is controlled indirectly through the auxiliary setup. This was accomplished by recording interference patterns between a weak diffracted beam copropagating with the XUV radiation and the probe beam. Figure~\ref{Fig_MainFringes} shows that the natural long-term drift without active stabilization (a) is essentially switched off upon activating the feedback loop (b), although a tiny residual drift of less than 300~\emph{as} per hour remains compared to the locking point, marked with red dashed line. Notably, this drift is absent, when scanning over 2.5 optical cycles during a three-hour long measurement (Fig.~\ref{Fig_MainFringes}(c)).

\subsection{Benchmark results}\label{subsect_ReMi_test}
Exceptional stability is vital for coincidence spectroscopy experiments, which involve analysis of particle behavior in a multidimensional---temporal, angular and energy resolved---manner. Additionally, the average ionization rate must not exceed the repetition rate of the laser source in order to maintain coincidence between photoions and photoelectrons (PEs). This constraint imposes a limitation on the acquisition rate.\par
As a proof-of-principle experiment, we performed an angle-resolved study of two-photon two-color (XUV+IR) ionization of argon gas using the RABBITT technique  \cite{Paul2001,Mairesse2003}. In this type of measurement, XUV harmonic radiation with photon energies of $(2q\pm1)\omega_0$ (where $q$ is an integer number and $\omega_0$ is the fundamental photon energy in atomic units) promote electrons of the target atom into the continuum. During the interaction with the weak field, the kinetic energy of these PEs shifts by $\pm\omega_0$  through the absorption or emission of an additional IR photon, resulting in the formation of SBs in the PE spectrum. Shifting the center of the IR envelope with respect to that of the XUV by a controlled $\tau$ relative delay, the SB signal $S_{2q}$ oscillates according to the relation:
\begin{equation}
S_{2q}(\tau) = S_0+A\mathrm{cos}[2\omega_0\tau+\phi],
\end{equation}
where $S_{0}$  is the baseline signal, $A$ is the amplitude, $\omega_0$ is the angular frequency of the IR field (equal to the photon energy in atomic units) and $\phi$ is a phase factor \cite{Dahlstrom2013,Isinger2019}.\par
\begin{figure*}
  \centering
  \includegraphics[trim=0.0cm 0.0cm 0.0cm 0.0cm,clip,width=0.65\linewidth]{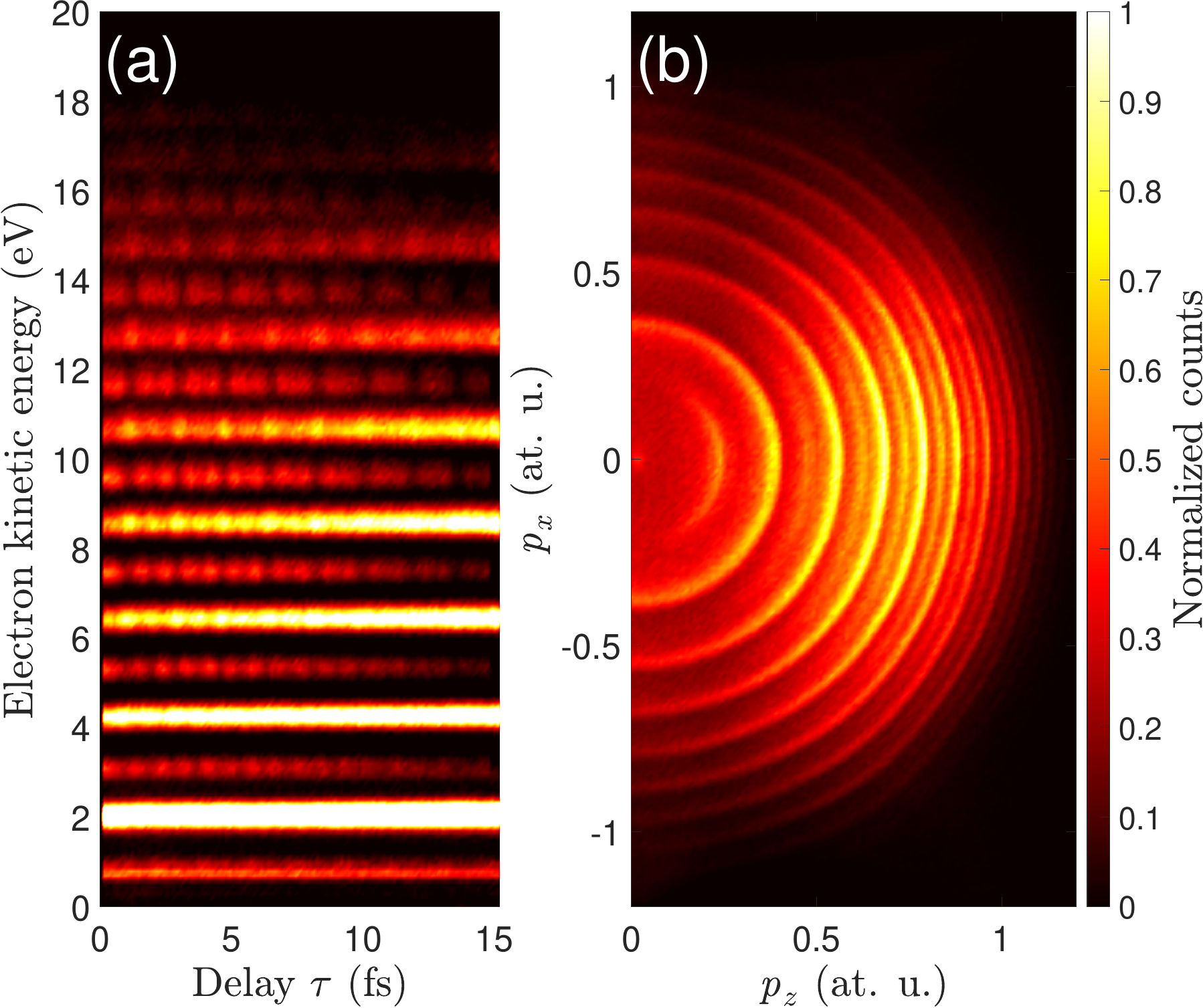}
  \caption{Time- and angularly resolved two-color two-photon photoionization of argon gas for the demonstration of long-term interferometric stability. (a): RABBITT trace recorded with a C-ReMi. (b): Delay-integrated momentum distribution of the same dataset in the plane of polarization. $p_z$ and $p_x$ mark momentum components parallel and perpendicular to the axis of the spectrometer, respectively.}
  \label{Fig_RABBITT}
\end{figure*}
Fig.~\ref{Fig_RABBITT}(a) demonstrates a RABBITT spectrogram recorded in argon target. Only PEs that are detected in coincidence with those Ar$^+$ ions, which originate from the supersonic, internally cold gas jet target of the C-ReMi were considered. False coincidence events were removed using momentum sum filters \cite{Prumper2007}. Under these conditions, the PE count rate using the 10~kHz repetition rate laser system was about 120~Hz, necessitating an acquisition time of about 14~hours to obtain sufficient particle statistics. Furthermore, in Fig.~\ref{Fig_RABBITT}(a), the emission direction of electrons were restricted to a small cone pointing towards the ion detector having an opening angle of 30$^\circ$ for maximizing the kinetic energy resolution. The trace reveals clear delay-dependent oscillations in the signal at various bands. The oscillation frequency increases with decreasing delay values and lower kinetic energies, due to the relatively intense probe field that opens pathways involving the transfer of more than one IR photon \cite{Swoboda2009}.\par
The corresponding PE momentum distribution in the polarization plane is presented in Fig.~\ref{Fig_RABBITT}(b). It exhibits a characteristic ringlike structure corresponding to PEs generated by the high harmonic comb, with SBs in between them. The angular distribution of the latter is noticeably different due to the different angular momentum channels participating in single- and two-color photoionization \cite{Peschel2022}. The linear polarization of the XUV radiation was slightly tilted with respect to the axis of the spectrometer ($z$) that is visible as a minor up-down asymmetry.\par
\begin{figure*}
  \centering
  \includegraphics[trim=0.0cm 0.0cm 0.0cm 0.0cm,clip,width=1.0\linewidth]{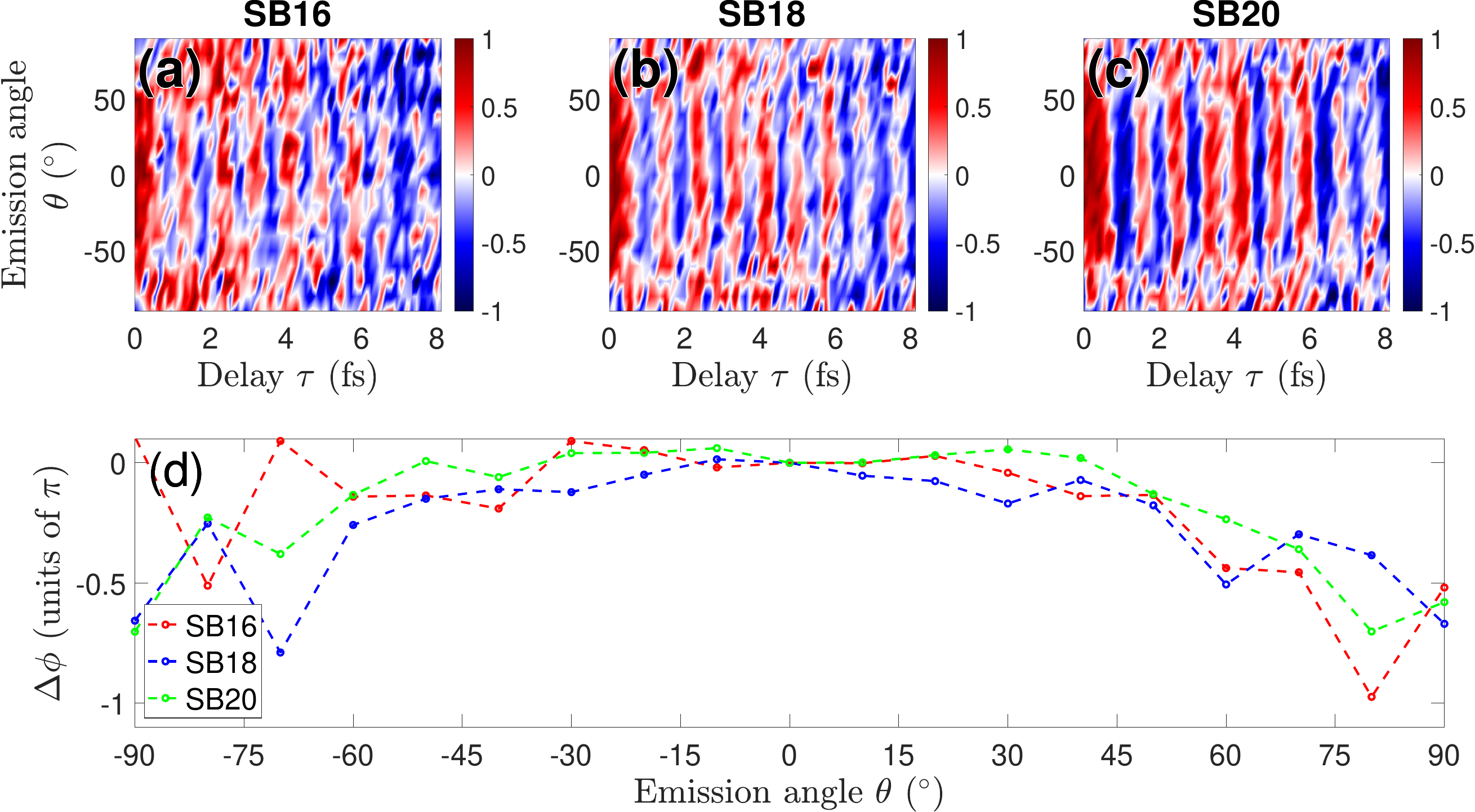}
  \caption{Angularly resolved oscillation of RABBITT SB signal corresponding to the (a) 16$^{\mathrm{th}}$, (b) 18$^{\mathrm{th}}$ and (c) 20$^{\mathrm{th}}$ harmonic of the fundamental photon energy. (d): Phase of the main oscillatory component as a function of emission angle for different SBs.}
  \label{Fig_AngularRABBITT}
\end{figure*}

Finally, Fig.~\ref{Fig_AngularRABBITT} shows the angularly resolved oscillation of different SBs corresponding to the $16^\mathrm{th}$, $18^\mathrm{th}$ and $20^\mathrm{th}$ harmonic of the fundamental photon energy as shown in panels (a), (b) and (c), respectively. The emission angle $\theta$ is defined as the angle between the direction of the PE emission and the spectrometer axis ($z$) in the polarization ($x$-$z$) plane. Clear oscillations are visible in the angularly resolved signal in case of all sidebands. These oscillations exhibit a drop in magnitude for the increasing absolute value of the emission angle, manifested as an increase in noise in the signal normalized for the total count rate at each $\theta$. Simultaneously, a phase shift in the sideband phase ($\Delta\phi$, where $\Delta\phi$ = $\phi(\theta)$-$\phi(0^\circ$)) exhibits a drastic, close to $\pi$ change (Fig.~\ref{Fig_AngularRABBITT}(d)). Although statistics in this particular measurement is insufficient to extract the photon energy dependence of the ratio and timing of individual partial waves contributing to the phase jump, such phenomenon has recently been studied theoretically \cite{Kheifets2023} and, for a single photon energy, experimentally in the form of an atomic level partial wave meter \cite{Jiang2022}. This reveals an exciting new application of attosecond coincidence metrology, a potential target for the system described in this work.
\section{Conclusion}\label{sect_conclusion}
\begin{hyphenrules}{nohyphenation}
We achieved long-term, sub-optical-cycle stability in a versatile, noncollinear interferometer with 15~m arm lengths and notably different wavelengths (XUV and IR) in the two arms of our system. This setup enables a fast feedback rate by using polarization-optical homodyne detection to monitor variations in the optical delay, while reflective and broadband optics ensure compatibility with ultrashort, few-cycle laser pulses. Additionally, this system can be integrated as a complementary setup to the main pump--probe interferometer without requiring significant modifications on the original layout and at moderate cost. Since optics subjected to high heat loads are common to both the XUV--IR and auxiliary interferometers, this approach is particularly well-suited for managing temporal instabilities caused by high-power driving lasers. It also offers scalability to higher laser powers and repetition rates, such as those provided by the HR-1 (100~kHz, 100~W) and HR-2 (100~kHz, 400~W) laser systems at ELI ALPS \cite{Charalambidis2017,Kuehn2017,Shirozhan2024}. Demonstrated through an angularly and time-resolved benchmark experiment on two-photon two-color photoionization, this day-long stability enables the study of light--matter interactions in a kinematically and temporally precise manner, opening up new possibilities for attosecond science.
\end{hyphenrules}






\section*{Acknowledgement}
The ELI ALPS project (GINOP-2.3.6-15-2015-00001) is supported by the European Union and co-financed by the European Regional Development Fund. B. M. is supported by the Janos Bolyai Research Scholarship of the Hungarian Academy of Sciences.

\section*{Disclosures}
The authors declare no conflicts of interest.

\bibliography{References}
\end{document}